\documentclass[12pt]{article}
\usepackage{amsmath,amsfonts,amssymb,bbm}
\usepackage{tensor}
\usepackage[T1]{fontenc}
\usepackage[utf8]{inputenc}
\usepackage{lmodern}
\usepackage{authblk}
\usepackage[disable]{todonotes}
\usepackage{color}
\usepackage{multicol}
\usepackage{cite}
\usepackage[margin=3cm]{geometry}
\usepackage{hyperref}

\newcommand*\xbar[1]{%
  \hbox{%
    \vbox{%
      \hrule height 0.5pt 
      \kern0.3ex
      \hbox{%
        \kern-0.0em
        \ensuremath{#1}%
        \kern-0.0em
      }%
    }%
  }%
}

\newcommand\email[1]{\thanks{\href{mailto:#1}{\nolinkurl{#1}}}}
\newcommand\td{\text{d}}
\newcommand\cO{{\cal O}}

\newcommand\cL{{\cal L}}
\newcommand\cG{{\cal G}}

\newcommand{\skyp}{{\cal I}^+}

\newcommand{\bz}{\bar{z}}
\newcommand{\bw}{\bar{w}}

\newcommand{\ffrac}[2]{\raisebox{.5pt}%
  {\footnotesize$\displaystyle\frac{#1}{#2}$}\kern1pt}

\newcommand{\ddl}[2]{\ffrac{\dd #1}{\dd #2}}

\newcommand{\vddl}[2]{{\ffrac{\delta #1}{\delta #2}}}

\newcommand{\dd}{\partial}
\renewcommand{\d}{\partial}

\def\cQ{\mathcal{Q}}
\allowdisplaybreaks[1]

\newcommand{\p}{\partial}
\def\n{\nabla}
\newcommand{\pb}{\p_{\bz}}
\newcommand{\be}{\begin{equation}}
\newcommand{\ee}{\end{equation}}
\newcommand{\bea}{\begin{eqnarray}}
\newcommand{\eea}{\end{eqnarray}}
\def\nn{\nonumber}
\def \ga {\gamma_{z\bz}}
\def \gai {\gamma_{z\bz}^{-1}}
\def \sg {\sqrt{-g}}

\def \gawi {\gamma_{w\bw}^{-1}}
\def \e{ \varepsilon}
\def \tit {\textit}
\def \g{g_{_{YM}}}

\def \bphi{\bar \phi}
\def \bPhi {\bar\Phi}


\author[a]{Pujian Mao\email{maopj@ihep.ac.cn}\,}
\author[b,c,d]{Jun-Bao Wu\email{junbao.wu@tju.edu.cn}\,}

\affil[a]{Institute of High Energy Physics and Theoretical Physics Center for Science Facilities, Chinese Academy of Sciences, 19B Yuquan Road, Beijing 100049, P.~R.~China}

\affil[b]{School of Science, Tianjin University, 92 Weijin Road, Tianjin 300072, P.~R.~China}
\affil[c]{School of Physics and Nuclear Energy Engineering, Beihang University, 37 Xueyuan Road, Beijing 100191, P.~R.~China}
\affil[d]{Center for High Energy Physics, Peking University, 5 Yiheyuan Road, Beijing 100871, P.~R.~China}

\title{\bf Note on asymptotic symmetries and soft gluon theorems\\}
\date{}

\begin{document}
 \maketitle
 \thispagestyle{empty}

\begin{abstract}
Recently, the leading soft gluon theorem with single soft emission was shown to be the Ward identity
of a two dimensional $\cG$-Kac-Moody symmetry. In this note, we show that the leading soft gluon theorem can be interpreted as the Ward identity for the asymptotic symmetries of non-Abelian gauge theory. We further argue that the sub-leading soft gluon theorem can follow from the same symmetry.
\end{abstract}


\section{Introduction}
\label{sec:introduction}
As the most perfect microscopic structures in the universe, scattering amplitudes have been undergoing a revolution in our understanding recently \cite{Dixon:2011xs}. One of the most remarkable discoveries is the on-shell recursion relation for tree amplitudes \cite{Britto:2004ap,Britto:2005fq} (also \cite{Cachazo:2013hca} from a different set-up). The perturbative S-matrix of massless excitations of the theory turns out to be much simpler than expected. The S-matrix is an object living at the boundary of space-time, so it is fair to ask whether this simplicity might be somehow connected
to the structure of null infinity which is accessible to the massless excitations.

Meanwhile, there has been renewed interest in asymptotic symmetries at null infinity. The novel advance is the promotion of the asymptotic symmetries to nontrivial symmetries of the quantum S-matrix \cite{Strominger:2013lka,Strominger:2013jfa}. The novel symmetries and the on-shell recursion relations converge at the understanding of the scattering of very low-energy massless quanta \textit{i.e.} soft theorems. On the one hand, the on-shell recursion relations provide a systematical way to derive soft factors at tree level \cite{Cachazo:2014fwa,Casali:2014xpa,Schwab:2014xua,Kalousios:2014uva,Zlotnikov:2014sva}. On the other hand, the soft theorems can be interpreted as Ward identities of asymptotic symmetries \cite{He:2014laa,He:2014cra} (see also \cite{Kapec:2014zla,Mohd:2014oja,Kapec:2015vwa,Campiglia:2015qka,Kapec:2015ena,Campiglia:2015kxa,Conde:2016csj,Conde:2016rom,Campoleoni:2017mbt} for further development).

With such connection in mind, one would expect that the emergence of one side would indicate the other. However, this is not always obvious. In practice, one can first associate certain soft theorem to the Ward identity of some unknown symmetries and then try to understand the nature of the unknown symmetries \cite{Kapec:2014opa,Lysov:2014csa,Campiglia:2014yka,Campiglia:2015yka,He:2015zea,Dumitrescu:2015fej,Lysov:2015jrs,Campiglia:2016jdj,Campiglia:2016hvg,Campiglia:2016efb,Kapec:2016jld,Campiglia:2017dpg}. In Yang-Mills theory, the soft gluon theorem \cite{Berends:1988zn} has been connected to the Ward identity of a holomorphic two dimensional $\cG$-Kac-Moody symmetry \cite{He:2015zea}. This new Kac-Moody symmetry is argued to be related to the asymptotic symmetries derived in \cite{Strominger:2013lka}. It would be definitely illuminating to show directly that the soft gluon theorem is just the Ward identity for the asymptotic symmetries of non-Abelian gauge theory following closely the line of \cite{He:2014laa,He:2014cra}. This is precisely what we will demonstrate in the following pages.

In this note, we study a matter coupled non-Abelian gauge theory. The reason for coupling matter is to have a consistent investigation into both the leading and sub-leading soft gluon theorem, while the tree-level sub-leading soft factor vanishes in pure Yang-Mills theory \cite{Bianchi:2014gla} under a certain prescription in which one solves the momentum conservation of the hard gluons according to the cyclic order in partial amplitudes (see also a supersymmetric extension \cite{Liu:2014vva}).
Our results show that the leading soft gluon theorem is nothing but the Ward identity for asymptotic symmetries of the matter coupled non-Abelian gauge theory. We also find that the sub-leading soft gluon theorem can follow from the same asymptotic symmetry responsible for the leading soft gluon theorem in a particular case where we consider only three-point vertices with one gluon coupling to two scalars.


This note is organized as follows. In the next section, we derive the general solutions of non-Abelian gauge theory and define the associated charges. Section \ref{sec:leading} is devoted to the connection of the leading soft gluon theorem and asymptotic symmetries. In section \ref{sec:subleading}, we argue that the sub-leading soft gluon theorem can follow the same symmetry in the scattering of $n$ scalars and one single soft gluon. We end this note with some discussions on future directions. There are also two appendixes in which we review the surface charge in non-Abelian gauge theories and the soft gluon theorems.

\section{4D non-Abelian gauge theory}
\label{sec:solution}

In this note, we will deal with theory living on the flat Minkowski spacetime in the retarded coordinates
\be\label{metric}
ds^2=\eta_{\mu\nu}\td x^\mu \td x^\nu=-\td u^2-2 \td u\td r+2r^2\ga \td z\td\bz,
\ee
where $\ga=\frac{2}{(1+z\bz)^2}$. This coordinate system will cover only future null infinity $\skyp$ at $r\to\infty$. The non-zero components of the Levi-Civit\`{a} connection are
\be
\Gamma^u_{z\bz}=r \ga,\;\;\;\;\Gamma^r_{z\bz}=-r\ga,\;\;\;\;\Gamma^z_{rz}=\frac{1}{r},\;\;\;\;\Gamma^z_{zz}=\p_z\ln\ga.
\ee

For simplicity, we consider a non-Abelian gauge theory with only scalar matter. The scalar fields are in the adjoint representation of the gauge group. The Lagrangian of the theory is
\be\label{lagrangian}
\cL=\frac14F_{\mu\nu}^aF^{a\mu\nu} + D_\mu\phi^a (D^\mu\phi^a)^\dag,
\ee
where the field-strength is defined by
\be
F_{\mu\nu}^a=\p_\mu A_\nu^a - \p_\nu A_\mu^a +i\g C^{abc}A^b_\mu A^c_\nu,
\ee
and $D_\mu$ is the covariant derivative associated to the local group $G$,
\be
D_\mu X^a=\n_\mu X^a + i\g C^{abc} A^b_\mu X^c.
\ee
The group structure constant $C_{abc}$ is totally antisymmetric and satisfies the Jacobi identity:
\be
C^{abc}C^{bde} + C^{abd}C^{bec} + C^{abe}C^{bcd}=0.
\ee
The equations of motion derived from the Lagrangian \eqref{lagrangian} are
\be\label{eom}
D_\mu F^{c\mu\nu}=\g J^{c\nu},\quad D_\mu D^\mu\phi^a=0,\quad D_\mu D^\mu\bphi^a=0,
\ee
where
\be\label{current}
J^{c\nu}=i C^{abc}(\phi^a D^\nu\bphi^b - \bphi^b D^\nu \phi^a).
\ee
The Lagrangian of the theory is invariant up to the total derivative under the gauge transformation
\be\label{gaugetransf}
\delta_\e A_\mu^a=D_\mu \e^a,\quad \delta_\e \phi^a=i\g C^{abc}\phi^b\e^c.
\ee

\subsection*{Asymptotic symmetries}
The asymptotic symmetry of 4 dimensional Yang-Mills theory was discovered recently in \cite{Strominger:2013lka} (see also \cite{Barnich:2013sxa,He:2015zea}). We will derive the asymptotic symmetries of non-Abelian gauge theory with adaption to our conventions.
The radial gauge
\be\label{condition}
A_r^a=0,
\ee
will be applied. 
The boundary condition of the rest components of the gauge fields is set to ensure the radiation flux is finite asymptotically:
\be\label{boundary}
A_u^a=\cO(r^{-1}),\quad A_{z(\bz)}^a=\cO(1),\quad \phi^a=\cO(r^{-1}),\quad \bphi^a=\cO(r^{-1}).
\ee
The conditions \eqref{condition} and \eqref{boundary} yield the residual gauge transformation parameter $\e^a(z,\bz)$. One can further check the symmetry algebra
\be
[\delta_{\e_{1}},\delta_{\e_{2}}]A_\mu^a=\delta_{[\e_{1},\e_2]}A_\mu^a,\quad [\delta_{\e_{1}},\delta_{\e_{2}}]\phi^a=\delta_{[\e_{1},\e_2]}\phi^a.
\ee

\subsection*{Asymptotic conserved charges}
The definition of charge in Yang-Mills theory is quite subtle in contrast with the Abelian case \tit{i.e.} Maxwell theory. The reason is that the field-strength is covariant under gauge transformation in a non-Abelian theory. However this issue was fixed in Yang-Mills theory asymptotically by Abbott and Deser\cite{Abbott:1982jh} (see also \cite{Barnich:2001jy} for a comprehensive analysis). We compute the asymptotic conserved charge for scalar matter coupled theory \eqref{lagrangian} in Appendix \ref{chargechapter} using the cohomological techniques \cite{Barnich:2001jy,Anderson:1996sc}. The charge is given by
\be\label{charge}
\cQ_{\e}=-\int\td z\td\bz\,r^2\,\ga\,\e^a \,F^{aur}.
\ee

\subsection*{Solution space}
In analogy with the charge associated to global symmetry, the charge \eqref{charge} is conserved on-shell asymptotically. Thus solving the equations of motion \eqref{eom} in series expansion around $\skyp$ is necessary for further investigation. As a gauge theory, not all the equations of motion are independent. They are connected off-shell by the Noether identity\footnote{The Noether identity of this theory is given in a more elegant way $D_\nu D_\mu F^{c\mu\nu} =0$. However, we find that \eqref{Bianchi} is more useful in practice.}
\be\begin{split}\label{Bianchi}
D_\nu\left[D_\mu F^{c\mu\nu}-\g J^{c\nu}\right]=i\g C^{abc}\left[\bphi^b D_\mu D^\mu\phi^a - \phi^a D_\mu D^\mu\bphi^b \right].
\end{split}\ee
It will be very helpful to arrange \eqref{eom} into
\begin{itemize}
\item hypersurface equations:
\be
D_\mu F^{a\mu u}=\g J^{au},
\ee
\item standard equations:
\be
D_\mu F^{a\mu z}=\g J^{az},\quad D_\mu F^{a\mu \bz}=\g J^{a\bz}, \quad D_\mu D^\mu\phi^a=0,\quad D_\mu D^\mu\bphi^a=0,
\ee
\item supplementary equations:
\be
D_\mu F^{a\mu r}=\g J^{ar}.
\ee
\end{itemize}
The advantage of such an arrangement is that \eqref{Bianchi} will be reduced to
\be
\p_r\left[\sg \big( D_\mu F^{a\mu r} - \g J^{ar}\big)\right]=0,
\ee
when the hypersurface equations and the standard equations are satisfied. Hence, one just needs to solve
\be
\sg (D_\mu F^{a\mu r}- \g J^{ar})=0
\ee
at order $\cO(r^0)$ and all the other orders (in $\frac1r$) will be zero automatically. That is the reason why they are called supplementary equations.

We start to solve the equations of motion \eqref{eom} with the hypersurface equations $D_\mu F^{a\mu u}=\g J^{au}$, from which one can solve out
\begin{multline}
A_u^a=\frac{A_u^{a(0)}(u,z,\bz)}{r}+\gai\int^\infty_r \frac{\td r'}{r'^2}\int^\infty_{r'} \td r'' \big[\p_z \p_{r''} A_{\bz}^a + \p_{\bz} \p_{r''} A_{z}^a \\
+i\g C^{abc}\big(A^b_z \p_{r''} A^c_{\bz} + A^b_{\bz} \p_{r''} A^c_{z} + {r''}^2\ga \phi^b \p_{r''} \bphi^c - {r''}^2\ga\bphi^c \p_{r''} \phi^b \big) \big],
\end{multline}
where $A_u^{a(0)}(u,z,\bz)$ are integration constants. Suppose that $A_{z}^a$, $A_{\bz}^a$, $\phi^a$, and $\bphi^a$ are given as initial data:
\be\begin{split}
&A_{z}^a=\sum\limits_{m=0}^\infty\frac{A_{z}^{a(m)}(u_0,z,\bz)}{r^{m}},\quad A_{\bz}^a=\sum\limits_{m=0}^\infty\frac{A_{\bz}^{a(m)}(u_0,z,\bz)}{r^{m}},\\
&\phi^a=\sum\limits_{m=0}^\infty\frac{\phi^{a(m)}(u_0,z,\bz)}{r^{m+1}},\quad \bphi^a=\sum\limits_{m=0}^\infty\frac{\bphi^{a(m)}(u_0,z,\bz)}{r^{m+1}}.
\end{split}\ee
$A_u^a$ are completely fixed up to the integration constants $A_u^{a(0)}(u,z,\bz)$. More precisely,
\be\label{Au}
A_u^a=\frac{A_u^{a(0)}}{r}-\frac{\gai A_u^{a(1)}}{2r^2}+\cO({r^{-3}}),
\ee
\be\label{Au1}
A_u^{a(1)}=\p_z A_{\bz}^{a(1)} + \p_{\bz} A_{z}^{a(1)} + i\g C^{abc}\big(A^{b(0)}_z A^{c(1)}_{\bz} - A^{b(1)}_z A^{c(0)}_{\bz} + \ga\, \phi^{b(0)} \bphi^{c(1)} - \ga\, \phi^{b(1)} \bphi^{c(0)}\big).
\ee

The standard equations will determine the time dependence of $A_{z}^{a(m)}$, $A_{\bz}^{a(m)}$, $\phi^{a(m)}$, and $\bphi^{a(m)}$ recursively.
In particular,
\begin{multline}\label{puAz1}
2\p_u A_z^{a(1)}=\p_z A^{a(0)}_u + \gai \p_z \big(\p_z A^{a(0)}_{\bz} - \p_{\bz} A^{a(0)}_{z}\big ) + i\g C^{abc}\big(\phi^{b(0)}\p_z\bphi^{c(0)} - \bphi^{c(0)} \p_z \phi^{b(0)}\big)\\
+ i \g \gai C^{abc} \p_z\big(A^{b(0)}_z A^{c(0)}_{\bz} \big) + i \g \gai C^{abc} A^{b(0)}_z \big(\p_z A^{c(0)}_{\bz} - \p_{\bz} A^{c(0)}_{z}\big)\\
- i \g C^{abc} A^{b(0)}_u A^{c(0)}_z + \gai \g^2 C^{abc}C^{dce} A^{b(0)}_z  A^{d(0)}_z  A^{e(0)}_{\bz}\\
 + \g^2 C^{abc}C^{dce} A^{d(0)}_z \big(\phi^{b(0)}\bphi^{e(0)} + \phi^{e(0)}\bphi^{b(0)}\big),
\end{multline}
\begin{multline}\label{puAzb1}
2\p_u A_{\bz}^{a(1)}=\p_{\bz} A^{a(0)}_u - \gai \p_{\bz} \big(\p_z A^{a(0)}_{\bz} - \p_{\bz} A^{a(0)}_{z}\big ) + i\g C^{abc}\big(\phi^{b(0)}\p_{\bz}\bphi^{c(0)} - \bphi^{c(0)} \p_{\bz} \phi^{b(0)}\big)\\
- i \g \gai C^{abc} \p_{\bz}\big(A^{b(0)}_z A^{c(0)}_{\bz} \big) - i \g \gai C^{abc} A^{b(0)}_{\bz} \big(\p_z A^{c(0)}_{\bz} - \p_{\bz} A^{c(0)}_{z}\big)\\
- i \g C^{abc} A^{b(0)}_u A^{c(0)}_{\bz} + \gai \g^2 C^{abc}C^{dce} A^{b(0)}_{\bz}  A^{d(0)}_{\bz}  A^{e(0)}_{z}\\
 + \g^2 C^{abc}C^{dce} A^{d(0)}_{\bz} \big(\phi^{b(0)}\bphi^{e(0)} + \phi^{e(0)}\bphi^{b(0)}\big),
\end{multline}
\begin{multline}\label{puphi1}
2\p_u \phi^{a(1)}=-2\gai\p_z\p_{\bz}\phi^{a(0)} - i\gai \g C^{abc}\big(\p_z A^{b(0)}_{\bz}\phi^{c(0)} + \p_{\bz} A^{b(0)}_{z} \phi^{c(0)}\big)\\
- i \g C^{abc}A^{b(0)}_u \phi^{c(0)} -2 i\gai \g C^{abc}\big(  A^{b(0)}_{\bz}\p_z \phi^{c(0)} +  A^{b(0)}_{z}\p_{\bz}\phi^{c(0)}\big) \\
 +\gai \g^2 C^{bec}C^{dca}\big(A_z^{b(0)}A^{d(0)}_{\bz}+A_z^{d(0)}A^{b(0)}_{\bz}\big)\phi^{e(0)},
\end{multline}
\begin{multline}\label{pubphi1}
2\p_u \bphi^{a(1)}=-2\gai\p_z\p_{\bz}\bphi^{a(0)} - i\gai \g C^{abc}\big(\p_z A^{b(0)}_{\bz}\bphi^{c(0)} + \p_{\bz} A^{b(0)}_{z} \bphi^{c(0)}\big)\\
 - i \g C^{abc}A^{b(0)}_u \bphi^{c(0)} - 2 i\gai \g C^{abc}\big(  A^{b(0)}_{\bz}\p_z \bphi^{c(0)} +  A^{b(0)}_{z}\p_{\bz}\bphi^{c(0)}\big) \\
 + \gai \g^2 C^{bec}C^{dca}\big(A_z^{b(0)}A^{d(0)}_{\bz}+A_z^{d(0)}A^{b(0)}_{\bz}\big)\bphi^{e(0)}.
\end{multline}
However there is no constraint on the leading terms $A_z^{a(0)}$, $A_{\bz}^{a(0)}$, $\phi^{a(0)}$, and $\bphi^{a(0)}$. We would refer to $\p_u A_z^{a(0)}$, $\p_u A_{\bz}^{a(0)}$, $\p_u \phi^{a(0)}$, and $\p_u \bphi^{a(0)}$ as \tit{news} functions following the terminology of \cite{Bondi:1962px,Sachs:1962wk,Barnich:2010eb}. The appearance of \tit{news} functions reflects the local propagating degree of freedom.

In the end, the supplementary equations
$D_\mu F^{a\mu r}=\g J^{ar}$ control the time evolution of the integration constants as
\begin{multline}\label{puAu0}
\p_u A_u^{a(0)}=\gai \p_u\big(\p_z A_{\bz}^{a(0)} + \p_{\bz} A_z^{a(0)}\big) + i \gai \g C^{abc}\big(A_z^{b(0)}\p_u A_{\bz}^{c(0)} -  A_{\bz}^{c(0)}\p_u A_z^{b(0)}\big)\\
+ i \g C^{abc}\big(\phi^{b(0)}\p_u \bphi^{c(0)} -  \bphi^{c(0)}\p_u \phi^{b(0)}\big).
\end{multline}

We summarize as follows: to specify one solution to the non-Abelian gauge system \eqref{lagrangian} with the conditions \eqref{condition} and \eqref{boundary}, one must specify the data $A_{z}^{a(m)}(u_0,z,\bz)$, $A_{\bz}^{a(m)}(u_0,z,\bz)$ $\phi^{a(m)}(u_0,z,\bz)$, $\bphi^{a(m)}(u_0,z,\bz)$ ($m \geq 1$) and $A_u^{a(0)}(u_0,z,\bz)$ at the initial time $u_0$, and specify the functions $A_z^{a(0)}(u,z,\bz)$, $A_{\bz}^{a(0)}(u,z,\bz)$, $\phi^{a(0)}(u,z,\bz)$ and $\bphi^{a(0)}(u,z,\bz)$ at any time on $\skyp$.

\section{Leading soft gluon theorem}
\label{sec:leading}

The soft gluon theorem \cite{Berends:1988zn,Casali:2014xpa} is a relation connecting an $n+1$ particle scattering where the extra particle is a soft gluon (with very low energy) to an $n$ particle scattering. In color-ordered amplitude form, it can be put in the following way:
\begin{equation}
\label{subsofth}
	M_{n+1}^{\pm}=\left(S_{\pm}^{(0)}+S_{\pm}^{(1)}\right)M_{n}+\cO\left(\omega\right), \quad \omega\rightarrow 0,
\end{equation}
where $\pm$ and $\omega$ denote the helicity and the energy of the extra soft gluon, respectively. On the right-hand side, $S_{\pm}^{(0)}$ and $S_{\pm}^{(1)}$ are the so-called leading and sub-leading (in $\omega$) soft factors. We revisit the soft gluon theorem in Appendix \ref{soft theorem}. The soft factors are given explicitly by equations \eqref{leadingfactor} and \eqref{subleadingfactor}. The relation \eqref{subsofth} is universal in the sense that it is valid for any type of matter minimally coupled at tree level\footnote{We will avoid commenting on loop corrections to soft gluon theorems in this work. Those relevant issues were well-studied, for instance, in \cite{Bern:1998sc,Bern:1999ry,Kosower:1999rx,Kosower:2003cz,Bern:2014oka,He:2014bga,Bern:2014vva}.}.

To connect soft theorems with asymptotic symmetries, one needs to recall the Ward identity of a spontaneously broken symmetry. In S-matrix language, a symmetry is just a relation between matrix elements $\langle\rm{out'}|\rm{in'}\rangle=\langle\rm{out}|\rm{in}\rangle$, where the \textit{in} and \textit{out} states are transformed as $|\rm{in'}\rangle=U^{\textrm{in}}|\rm{in}\rangle$ and $|\rm{out'}\rangle=U^{\textrm{out}}|\rm{out}\rangle$.  If a conserved charge $Q$ can be associated to the symmetry, the Ward identity becomes
\begin{equation}
\label{Ward}
	\langle\rm{out}|Q^{\textrm{out}}-Q^{\textrm{in}}|\rm{in}\rangle=0.
\end{equation}
The charge for a spontaneously broken symmetry must act non-linearly on the states, otherwise it will annihilate the vacuum. In principle, it can be split into a linear piece that annihilates the vacuum and a non-linear piece that does not annihilate the vacuum
\be
Q=Q_{\rm{L}}+Q_{\rm{NL}}.
\ee
The Ward identity for a broken charge is
\begin{equation}
\label{WardNL}
	\langle\rm{out}|Q_{\rm{NL}}^{\textrm{out}}-Q^{\textrm{in}}_{\rm{NL}}|\rm{in}\rangle=
	-\langle\rm{out}|Q^{\textrm{out}}_{\rm{L}}-Q^{\textrm{in}}_{\rm{L}}|\rm{in}\rangle.
\end{equation}
If $Q_{\textrm{NL}}$ creates zero-momentum Goldstone bosons, equation \eqref{WardNL} looks very much like the soft theorem \eqref{subsofth}. We will show precisely that the broken symmetry responsible for the soft gluon theorem is the asymptotic symmetry preserving the prescribed conditions \eqref{condition} and \eqref{boundary} with the associated charge \eqref{charge} in this section.

It is important to point out that we will only deal with the \tit{out} part of the states in this note. The analysis of the \tit{in} part can be performed analogously, up to an anti-podal identification. These details have been laid out for other theories in \cite{He:2014laa,He:2014cra} (see also \cite{He:2015zea} for the Yang-Mills case), which can be easily extended to the non-Abelian gauge theory that we are exploring.

Following \cite{He:2014laa,He:2014cra}, we evaluate the charge \eqref{charge} at the far past of the future null infinity $\skyp_-$. The leading piece of the charge is
\be\begin{split}
\cQ^{(0)}&=\int_{\skyp_-} \td z \td\bz \,\ga\, \e^a A_u^{a(0)}=\int_{\skyp}\td u \td z\td\bz\,\ga\, \e^a\p_u A_u^{a(0)}\\
&=\int_{\skyp}\td u \td z\td\bz\, \e^a \p_u\big(\p_z A_{\bz}^{a(0)} + \p_{\bz} A_z^{a(0)}\big) + i \e^a \g C^{abc}\big(A_z^{b(0)}\p_u A_{\bz}^{c(0)} -  A_{\bz}^{c(0)}\p_u A_z^{b(0)}\big)\\
&\quad +\ga\, i \g \e^a C^{abc}\big(\phi^{b(0)}\p_u \bphi^{c(0)} -  \bphi^{c(0)}\p_u \phi^{b(0)}\big).
\end{split}\ee
The charge splits into linear and non-linear pieces\footnote{One should not confuse with the fact that the linear piece of the charge consists of nonlinear fields while the nonlinear piece of the charge consists of linear fields.}
\begin{align}
\cQ^{(0)}_{\textrm{L}}=\int_{\skyp}\td u\td z\td\bz\,\g\, i \e^a \g C^{abc}&\big(A_z^{b(0)}\p_u A_{\bz}^{c(0)} -  A_{\bz}^{c(0)}\p_u A_z^{b(0)}\big)\nn\\
& + \ga\,i \g \e^a C^{abc}\big(\phi^{b(0)}\p_u \bphi^{c(0)} -  \bphi^{c(0)}\p_u \phi^{b(0)}\big),\label{Q0L}\\
\cQ^{(0)}_{\textrm{NL}}=\int_{\skyp}\td u\td z\td\bz\,\e^a \p_u\big(\p_z A_{\bz}^{a(0)}+ &\p_{\bz} A_z^{a(0)}\big).\label{Q0NL}
\end{align}

For the \tit{news} fields, one needs to perform a stationary-phase approximation of the gauge field mode expansion:
\begin{equation}\label{stationary}
	A^{a(0)}_{z(\bz)}(x)=-\frac{i\g}{8\pi^2}\frac{\sqrt{2}}{1+z\bz}\int_0^\infty\td \omega_q\left[
\mathfrak{a}_{+(-)}^a(\omega_q\hat{x})\,e^{-i\omega_q u}-\mathfrak{a}^{a\dagger}_{-(+)}(\omega_q\hat{x})\,e^{i\omega_q u}\right],
\end{equation}
where the creation and annihilation operators satisfy the standard commutation relations. Then, using the Fourier relation,
\begin{equation}
F(u)=\int_{-\infty}^{\infty}\td\omega\,e^{i\omega u}\tilde{F}(\omega),\quad	\int_{-\infty}^{\infty}\td u\,\partial_uF(u)=2\pi i\lim_{\omega\to0}\left[\omega\tilde{F}(\omega)\right],
\end{equation}
and a special choice of the gauge parameter
\begin{equation}\label{e}
\varepsilon^a(z,\bz)=\frac{1}{w-z},
\end{equation}
which leads to $\partial_{\bz}\varepsilon^a=-2\pi\delta^2(z-w)$, we obtain for the non-linear pieces of the charge\footnote{
We are just keeping the anti-holomorphic parts of the the charges \eqref{Q0NL}, meaning those containing only $\partial_{\bz}\varepsilon$. Consequently, one has to split \eqref{Q0L} via $\cQ^{(0)}_{\textrm{L}}\to\frac12\cQ^{(0)}_{\textrm{L}}+\frac12\cQ^{(0)}_{\textrm{L}}$. Otherwise one has to
introduce some extra factors of 2, which arise from a proper treatment of the radiative modes \cite{He:2014cra,Mohd:2014oja}. The same treatment will be applied for the analysis of the sub-leading soft theorem in the next section.}
\begin{align}\label{NLleading}
	\langle\textrm{out}|\cQ^{(0)}_{\textrm{NL}}|\textrm{in}\rangle&=\frac{1}{4}\frac{\sqrt{2}\g}{1+|w|^2}\lim_{\omega_q\to0}
	\langle\textrm{out}|\omega_q\,\mathfrak{a}^a_+(q)|\textrm{in}\rangle_{i_1 \cdots i_n},
\end{align}
where $a$ and $i_k\,(k=1,2,\cdots n)$ are color indices.

In order to obtain the proper action of the linear pieces of the charge on the \textit{out} states, one has to define canonical commutation relations at infinity. The non-vanishing ones are
\begin{align}
  [A^{a(0)}_{\bz}(u,z,\bz),A_z^{b(0)}(u',w_k,\bw_k)]&=\frac14 \,\delta^{ab}\, \Theta(u'-u)\delta^2(z-w_k),\label{Acommutator} \\
  [\xbar\phi^{a(0)}(u,z,\bz),\phi^{b(0)}(u',w_k,\bw_k)]&=\frac14 \,\delta_{ab}\gai\, \Theta(u'-u)\delta^2(z-w_k).\label{phicommutator}
\end{align}
We define the Fourier modes as
\be\begin{split}
&A_{E_k}^d(w_k,\bw_k)=\int \td u' e^{iE_k u'} A^{d(0)}_z(u',w_k,\bw_k),\\
&\phi_{E_k}^d(w_k,\bw_k)=\int \td u' e^{iE_k u'} \phi^{d(0)}(u',w_k,\bw_k).
\end{split}\ee
The actions of the fields on the Fourier modes are
\begin{align}
&[A^{a(0)}_{\bz},A_{E_k}^d(w_k,\bw_k)]=i\frac14\delta^{ad} \delta^2(z-w_k) \dfrac{e^{iE_k u}}{E_k},\\
&[\bphi^{a(0)},\phi_{E_k}^d(w_k,\bw_k))]=i\frac14\delta^{ad} \delta^2(z-w_k) \dfrac{e^{iE_k u}}{E_k},
\end{align}
which yield
\begin{align}
&[\cQ^{(0)}_{\textrm{L}},A_{E_k}^d(w_k,\bw_k)]=-\frac{i\g C_k^{abd}}{4(w-w_k)} A^b_{E_k}(w_k,\bw_k),\\
&[\cQ^{(0)}_{\textrm{L}},\phi_{E_k}^d(w_k,\bw_k)]=-\frac{i\g C_k^{abd}}{4(w-w_k)} \phi^b_{E_k}(w_k,\bw_k),
\end{align}
where the special choice of $\e^a$ in \eqref{e} has been used. Hence
\be
\label{Q0Lbk}
  \langle\textrm{out}|Q^{(0)}_{\textrm{L}}|\textrm{in}\rangle=-\sum_{k=1}^n
  \frac{\g (T_k^a)_{i_kj_k}}{4(w-w_k)}\langle\textrm{out}|\textrm{in}\rangle_{i_1\cdots j_k \cdots i_n},
\ee
where $(T_k^a)_{i_kj_k}=-i C_k^{ai_kj_k}$ in adjoint representation. Inserting \eqref{Q0Lbk} and \eqref{NLleading} into the Ward identity \eqref{WardNL}, one should be able to reproduce the leading soft gluon theorem \eqref{soft0+} in asymptotic position space.

\section{Sub-leading soft gluon theorem}
\label{sec:subleading}

After the realization that the leading soft gluon theorem follows from the asymptotic symmetry of the non-Abelian gauge theory, whether
or not the sub-leading soft gluon theorem follows from new asymptotic symmetries becomes a very important question. In Abelian gauge theory and linearized gravity theory, the sub-leading soft theorems were shown \cite{Conde:2016csj,Conde:2016rom} to be equivalent to the Ward identities for the sub-leading pieces (in $\frac1r$) of the charge associated to the same asymptotic symmetry responsible for the leading soft theorem. It is very interesting to check whether this prescription can recover the sub-leading soft gluon theorem in our case.

The sub-leading piece of the charge \eqref{charge} is
\be\begin{split}\label{subcharge}
\cQ^{(1)}=&-\frac1r\int_{\skyp_-} \td z \td\bz \,\e^a A_u^{a(1)}\\
=&-\frac1r\int_{\skyp_-} \td z \td\bz \,\e^a\bigg[\p_z A_{\bz}^{a(1)} + \p_{\bz} A_{z}^{a(1)}\\
&+i\g C^{abc}\big(A^{b(0)}_z A^{c(1)}_{\bz} - A^{b(1)}_z A^{c(0)}_{\bz} + \ga\, \phi^{b(0)} \bphi^{c(1)} - \ga\, \phi^{b(1)} \bphi^{c(0)}\big)\bigg].
\end{split}\ee

The next step should be checking the action of the charge on the states using the stationary-phase approximation \eqref{stationary} and the commutation relations \eqref{Acommutator}-\eqref{phicommutator}. However, we find that this is very hard, if not impossible, to achieve due to the complicated expressions in \eqref{puAz1}-\eqref{pubphi1}. Another difficulty is from the translation of the sub-leading soft factor \eqref{subleadingfactor} to asymptotic position space. It is not yet clear how to translate the helicity terms in the angular momentum operator in the sub-leading soft factor to asymptotic position space. As a first step to check the proposal of \cite{Conde:2016csj,Conde:2016rom} in the non-Abelian case, we focus on a particular case for the linearized part of the theory. We make the change of variables
\be
A^a_\mu\to{\cal A}_\mu^a + \g a^a_\mu,\quad \phi^a\to \Phi^a + \g \varphi^a,
\ee
where $({\cal A}_\mu^a,\Phi^a)$ is a background solution of the equations of motion \eqref{eom}, and we consider the lowest non-trivial order of the theory \tit{i.e.} the quadratic order (in $\g$) in the Lagrangian. In terms of Feynman diagrams, we will only deal with three-point vertices in which one gluon couples to two scalars in this note. The gluon three-point vertices and more point vertex are left for future investigation. For simplicity, we choose the background solution of the gauge field ${\cal A}_\mu^a=0$. With such simplification, the effective currents $J^{c\mu}$ defined in \eqref{current} become globally conserved currents while equations \eqref{puAz1}, \eqref{puAzb1}, and \eqref{puAu0} reduce to
\begin{align}
\label{puAz1s}
&2\p_u a_z^{a(1)}=\p_z a^{a(0)}_u + \gai \p_z \big(\p_z a^{a(0)}_{\bz} - \p_{\bz} a^{a(0)}_{z}\big ) +  j_z^a,\\
\label{puAzb1s}
&2\p_u a_{\bz}^{a(1)}=\p_{\bz} a^{a(0)}_u - \gai \p_{\bz} \big(\p_z a^{a(0)}_{\bz} - \p_{\bz} a^{a(0)}_{z}\big ) +  j_{\bz}^a,\\
\label{puAu0s}
&\p_u a_u^{a(0)}=\gai \p_u\big(\p_z a_{\bz}^{a(0)} + \p_{\bz} a_z^{a(0)}\big) +   j_u^a,
\end{align}
where
\be\label{current3}\begin{split}
&j_z^a=i C^{abc}(\Phi^{b(0)}\p_z\bPhi^{c(0)} - \bPhi^{c(0)} \p_z \Phi^{b(0)}),\\
&j_{\bz}^a=i C^{abc}(\Phi^{b(0)}\p_{\bz}\bPhi^{c(0)} - \bPhi^{c(0)} \p_{\bz} \Phi^{b(0)}),\\
&j_u^a= i C^{abc}(\Phi^{b(0)}\p_u \bPhi^{c(0)} -  \bPhi^{c(0)}\p_u \Phi^{b(0)}).
\end{split}\ee
The sub-leading charge becomes
\be\label{drop}
\cQ^{(1)}=-\frac{\g}{r}\int_{\skyp_-} \td z\td\bz\, \e^a (\p_z a_{\bz}^{a(1)} + \p_{\bz} a_{z}^{a(1)}+ \ga j_r^a),
\ee
where
\be\label{currentr}
j_r^a=i C^{abc}( \Phi^{b(0)} \bPhi^{c(1)} - \Phi^{b(1)} \bPhi^{c(0)}).
\ee
One can recognize that $j^a_\mu$ are just the leading terms of the conserved current $J^a_\mu$. In \cite{Conde:2016csj}, it is very crucial to set the radial component of the current to zero by the ambiguities of a conserved current to adapt to the radial gauge condition\footnote{Such treatment is also needed in the gravity case \cite{Conde:2016rom}, in which the stress-energy tensor was modified.}. Applying the same reasoning, we will drop the $j^a_r$ term from the sub-leading charge \eqref{drop}.

Now the sub-leading charge can be arranged into
\be\begin{split}
\cQ^{(1)}&=-\frac{\g}{r}\int_{\skyp_-} \td z\td\bz\, \e^a (\p_z a_{\bz}^{a(1)} + \p_{\bz} a_{z}^{a(1)})=-\frac{\g}{r}\int_{\skyp}\td u \td z\td\bz\,\e^a \p_u(\p_z a_{\bz}^{a(1)} + \p_{\bz} a_{z}^{a(1)})\\
&=-\frac{\g}{r}\int_{\skyp}\td u \td z\td\bz\,\e^a\left[\p_u(u\p_z\p_{\bz} a^{a(0)}_u) - u \p_z \p_{\bz} \p_u a^{a(0)}_u+\frac12 (\p_z j_{\bz}^a+\p_{\bz}j_z^a)\right]\\
&=\frac{\g}{r}\int_{\skyp}\td u \td z\td\bz\,\e^a\bigg[u\p_z\p_{\bz}\big(\gai \p_u(\p_z a_{\bz}^{a(0)} + \p_{\bz} a_z^{a(0)})\big)\\
&\hspace{6.5cm}+  u\p_z\p_{\bz} j^a_u -\frac12 (\p_z j_{\bz}^a+\p_{\bz}j_z^a)\bigg],
\end{split}\ee
where we have used integration by parts several times and one can drop the boundary term
\be
\int_{\skyp} \td z \td\bz \td u\,\p_u\left(u\, \e^a\, \p_z\pb a^{a(0)}_u\right)=-\lim_{u\to\infty} u\, \int_{\skyp_-} \td z \td\bz \, \p_z\pb\e^a\, a^{a(0)}_u,
\ee
which is simply a consequence of\footnote{The $u$ factor in front should not be worrisome, as it is still combined with a $1/r$ factor.}
\be
\langle\textrm{out}|Q^{(0)}_{\e^a_\textrm{out}=\p_z\pb\e^a}|\textrm{in}\rangle=0.
\ee

The charge also splits into linear and non-linear pieces
\begin{align}
  \cQ^{(1)}_{\textrm{L}}&=\frac{\g}{2r}\int_{\skyp}\td u\td z\td\bz\, \left[u(\p_z\p_{\bz}\e^a + \p_{\bz}\p_z\e^a) j^a_u + \p_z\e^a j_{\bz}^a + \p_{\bz}\e^a j_z^a\right],\\
  \cQ^{(1)}_{\textrm{NL}}&=\frac{\g}{r}\int_{\skyp}\td u\td z\td\bz\,\left(-\p_z(\gai\p_z(u\p_u a_{\bz}^{a(0)}))\p_{\bz}\e^a - \p_{\bz}(\gai\p_{\bz}(u \p_u a_z^{a(0)}))\p_z\e^a \right).
\end{align}
By exploiting the stationary-phase approximation
\begin{equation}
a^{a(0)}_{z(\bz)}(x)=-\frac{i}{8\pi^2}\frac{\sqrt{2}}{1+z\bz}\int_0^\infty\td \omega_q\left[
\mathfrak{a}_{+(-)}^a(\omega_q\hat{x})\,e^{-i\omega_q u}-\mathfrak{a}^{a\dagger}_{-(+)}(\omega_q\hat{x})\,e^{i\omega_q u}\right],
\end{equation}
and the Fourier relation
\be
\int_{-\infty}^{\infty}\td u\,u\,\partial_u F(u)=-2\pi \lim_{\omega\to0}\left[\partial_{\omega}\left(\omega\tilde{F}(\omega)\right)\right],
\ee
one gets the action of the non-linear piece of the charge\footnote{The factor $\frac{\g}{r}$ has been omitted as it will be canceled in the Ward identity \eqref{WardNL}.}
\be\label{NLsubleading}
\langle\textrm{out}|Q^{(1)}_{\textrm{NL}}|\textrm{in}\rangle=
  -\frac{\sqrt{2}i}{4}\,\p_w\left[\gawi\p_w\left(\frac{1}{1+|w|^2}\lim_{\omega_q\to0}\!\partial_{\omega_q}
  \langle\textrm{out}|\omega_q\,\mathfrak{a}^a_-(q)|\textrm{in}\rangle_{i_1 \cdots i_n}\right)\right],
\ee
where the special choice of $\e^a$ in \eqref{e} has been inserted.

From the commutation relations
\begin{align}
[\xbar\Phi^{a(0)}(u,z,\bz),\Phi^{b(0)}(u',w_k,\bw_k)]=\frac14 \,\delta_{ab}\gai\, \Theta(u'-u)\delta^2(z-w_k),
\end{align}
the action of the linear piece of the charge is
\begin{multline}
[\cQ^{(1)}_{\textrm{L}},\Phi_{E_k}^d(w_k,\bw_k)]=-\frac{ i \pi C_k^{abd}}{2}\\
\left(\partial_w(\gamma_{w\bw}^{-1}\delta(w-w_k))\partial_{\omega_k}+\gamma_{w\bw}^{-1}\,\omega_k^{-1}\delta(w-w_k)\partial_{w_k}\right)\phi^b_{E_k}(w_k,\bw_k),
\end{multline}
where $\Phi_{E_k}^d(w_k,\bw_k)=\int \td u' e^{iE_k u'} \Phi^{d(0)}(u',w_k,\bw_k)$ is the Fourier mode of the scalar field.
Hence
\begin{multline}
\label{Q1Lbk}
\langle\textrm{out}|Q^{(1)}_{\textrm{L}}|\textrm{in}\rangle=-\sum_{k=1}^n\frac{\pi (T_k^a)_{i_kj_k}}{2}\\
\left(\partial_w(\gamma_{w\bw}^{-1}\delta(w-w_k))\partial_{\omega_k}+\gamma_{w\bw}^{-1}\,\omega_k^{-1}\delta(w-w_k)\partial_{w_k}\right)\langle\textrm{out}|\textrm{in}\rangle_{i_1\cdots j_k \cdots i_n}.
\end{multline}
To recover the sub-leading soft gluon theorem \eqref{soft1+}, one just needs to insert \eqref{NLsubleading} and \eqref{Q1Lbk} into the Ward identity \eqref{WardNL}, and the relation
\begin{multline}
\p_w\left\{\gawi\p_w\left[\frac{1}{1+|w|^2}\left(\frac{1+w\,\bw_k}{\bw-\bw_k}\partial_{\omega_k}+
\frac{(1+|w_k|^2)(w-w_k)}{\omega_k(\bw-\bw_k)}\partial_{w_k}\right)\right]\right\}\\
=-2\pi\,\gamma_{w\bw}\left(\partial_w\left(\gamma_{w\bw}^{-1}\delta(w-w_k)\right)\partial_{\omega_k}
+\gamma_{w\bw}^{-1}\,\omega_k^{-1}\delta(w-w_k)\partial_{w_k}\right)
\end{multline}
needs to be utilized.

\section{Discussions}
Based on recent proposals in QED and gravitational theory that soft theorems can be interpreted as Ward identities of certain asymptotic symmetries \cite{He:2014laa,He:2014cra}, we have shown precisely the equivalence of the leading soft gluon theorem and the Ward identity for the asymptotic symmetry of non-Abelian gauge theory. Following the prescription in \cite{Conde:2016csj,Conde:2016rom}, we also show that the sub-leading soft gluon theorem can actually follow from the same symmetry in a simpler set-up for which we consider only vertices where a gluon couples to two scalars.

There are several open questions which need detailed investigation in the future. An immediate one is whether we can recover the sub-leading soft gluon theorem from the Ward identity of the sub-leading charge, including all vertices of the non-Abelian gauge theory. There are technical difficulties from the sub-leading charge and the sub-leading soft factor with helicity terms. Nevertheless, we believe that this problem can be fixed with very careful analysis since the prescription in \cite{Conde:2016csj,Conde:2016rom} presents a remarkably consistent matching of two expansions and it also partially succeeds in deriving the sub-leading gluon theorem.

Another interesting question comes from the multiple soft emissions \cite{Berends:1988zn} (see also \cite{Cachazo:2015ksa} for recent development). In such a case, the soft factor does not have a simple factorization even at the leading order; namely, the soft factor is related to the order of the gluons taken to be soft. Such a phenomenon is related to the structure of the symmetry group controlling the soft theorem \cite{ArkaniHamed:2008gz}. It would be a very enlightening investigation to check the possible connection of the soft theorems with multiple soft emissions and asymptotic symmetries.


\section*{Acknowledgments}

The authors thank Glenn Barnich and Eduardo Conde for useful discussions. This work is supported in part by the National Natural Science Foundation of China Grant No. 11575202.

\appendix

\section{Surface charge in non-Abelian gauge theory}\label{chargechapter}

We will derive the surface charge for non-Abelian gauge theory \eqref{lagrangian} using the cohomological technique \cite{Barnich:2001jy,Anderson:1996sc}. Such an approach is based on the equivalence classes of the Lagrangian up to total divergences.

We use $\phi^i$ to denote the fields in general in the variational principle. In an $n$ dimensional spacetime, the theory is defined by the
Lagrangian $L={\mathcal L}\, \td^nx$, where $\td^nx$ is the spacetime volume form. The notation
\begin{equation}
(d^{n-p}x)_{\mu_1\dots\mu_p}=\frac{1}{p!(n-p)!}
\epsilon_{\mu_1\dots\mu_p\mu_{p+1}\dots
  \mu_n}\td x^{\mu_{p+1}}\dots \td x^{\mu_n}
\end{equation}
will be used in this appendix where $\epsilon_{\mu_1\dots\mu_n}$ is completely antisymmetric. The Lagrangian is invariant up to total divergences under the non trivial gauge transformations
\be
\delta_f \phi^i=R^i_\alpha(f^\alpha).
\ee

The construction of the $n-2$ forms can be summarized as follows: for any $f^\alpha$, standard integrations by parts lead to
\begin{equation}
  R^i_\alpha(f^\alpha)\vddl{L}{\phi^i}=f^\alpha
  R^{+i}_\alpha(\vddl{L}{\phi^i})
  +\td_H S_f,
\label{eq:1}
\end{equation}
for some $n-1$ form
\begin{equation}
S_f=S^{i\mu}_\alpha(\ddl{}{\td x^\mu}\vddl{L}{\phi^i},f^\alpha) \label{eq:4}
\end{equation}
vanishing on-shell. The $n-2$ form will be obtained by acting the contracting homotopy operator $\rho_H$ for the horizontal differential of the variational bi-complex \cite{Andersonbook}
\begin{equation}
  \label{eq:5}
  \{\td_H,\rho_H\}\omega^p=\omega^p\ {\rm for}\ p<n
\end{equation}
on $S_f$,
\begin{equation}
  \label{eq:7}
  k_f=\rho_H S_f.
\end{equation}

For particular gauge transformations that satisfy $R^i_\alpha(f^\alpha)=0$ (we refer to them as reducibility parameters), \eqref{eq:1} leads to $\td_H S_f=0$. Hence \eqref{eq:5} reduces to
\begin{equation}
  \label{eq:8}
  \td_H k_{f}=S_f\approx 0.
\end{equation}

However, the reducibility parameter does not exist in general in many theories (\tit{e.g.} Yang-Mills theory and Einstein gravity). Normally, one would linearize such theories around a background solution $\bar\phi^i$ for deriving the surface charges. For instance, using the linearized theory at infinity with prescribed asymptotics to define conservation laws in general relativity is well explained in \cite{Misner:1974qy}. In the linearized theory, the gauge transformations are denoted by $\delta_\epsilon\varphi^i=R^i_\alpha[\varphi^i,\bar\phi](\epsilon^\alpha)$.
It has been shown \cite{Barnich:2003xg} that one can obtain the $n-2$ forms for the linearized theory from the full theory through
\begin{equation}
  \label{eq:9}
  k_f[\delta\phi,\phi]=k^{\mu\nu}_f(\td^{n-2}x)_{\mu\nu}=\frac{|\lambda|+1}{|\lambda|+2}
\partial_{(\lambda)}[\delta\phi^i\vddl{}{\phi^i_{((\lambda)\nu)}}\ddl{}{\td x^\nu}
S_f],
\end{equation}
by the following arrangements: $f$ replaced by the reducibility parameters of the linearized theory, $\phi^i$ replaced by the background solution $\bar\phi^i$ and $\delta\phi^i$ replaced by any solution $\bar\varphi^i$ of the linearized theory. In particular, for $S_f$ up to second order derivatives, one can reduce \eqref{eq:9} to
\begin{equation}
  \label{eq:64}
  k_f[\delta\phi,\phi]=\frac{1}{2}\delta\phi^i\vddl{}{\phi^i_\nu}\ddl{}{\td x^\nu}
  S_f+\frac{2}{3}\d_\sigma[\delta\phi^i\vddl{}{\phi^i_{\nu\sigma}}\ddl{}{\td x^\nu}
    S_f].
\end{equation}

Associating conserved charges to asymptotic symmetries is a somewhat tricky question \cite{Glenn}. In practice, one can take a more pragmatic step \cite{Barnich:2001jy} that lies in using the formula for the $n-2$ forms above, but substituting asymptotic reducibility parameters and asymptotic solutions determined by the fall-off conditions instead of the exact ones from linearized theory. Consequently, the $n-2$ forms are in general non-integrable. 
This is precisely what happens in \cite{Wald:1999wa,Barnich:2011mi,Barnich:2013axa,Barnich:2015jua,Barnich:2016rwk}. We will follow the same strategy for getting the surface charges associated to asymptotic symmetries in non-Abelian gauge theory.

Applying \eqref{eq:1} to the Lagrangian \eqref{lagrangian} with the precise gauge transformations \eqref{gaugetransf}, the $n-1$ form is obtained as
\be
S_f[\delta\phi,\phi]=-\n_\nu\left(f^a F^{a\mu\nu}\right)(\td^{n-1}x)_\mu.
\ee
Using \eqref{eq:64}, one finds the $n-2$ form is actually integrable and given by
\be
k_f[\delta\phi,\phi]=-\delta\left(f^a F^{a\mu\nu}\right)(\td^{n-2}x)_{\mu\nu}.
\ee

\section{Leading and sub-leading soft gluon theorems}\label{soft theorem}
For completeness, we review the leading and sub-leading soft gluon theorems. We have set all of the momenta outgoing by using crossing symmetry. The discussion of the sub-leading soft limit follows closely the one in~\cite{Bern:2014vva}. We denote the amplitude with $n$ particles as $A(p_1, \cdots, p_n)_{i_1 \cdots i_n}$ and the amplitude with the extra one gluon as $\epsilon_{q\mu} A^\mu(p_1, \cdots, p_n, q)_{i_1 \cdots i_n a}$ where $i_k$ and $a$ are color indices. We have
\begin{multline}
A^\mu(p_1, \cdots, p_n, q)_{i_1 \cdots i_n a}=\sum_{k=1}^n \frac{-i}{(q+p_k)^2} i (q+2p_k)^\mu (T^a_k)_{i_kj_k}A(p_1, \cdots, p_k+q, \cdots, p_n)_{i_1\cdots j_k \cdots i_n}\\
+N^\mu (p_1, \cdots, p_n, q)_{i_1\cdots i_n a},
\end{multline}
where $(T^a_k)_{i_kj_k}\equiv (T^a_{R_k})_{i_kj_k}$. Gauge invariance yields
\bea
0&=&q_\mu A^\mu(p_1, \cdots, p_n, q)_{i_1 \cdots i_n a}\nn\\
&=&\sum_{k=1}^n (T^a_k)_{i_kj_k}(1+q_\mu\frac{\partial}{\partial p_{k\mu}})A(p_1, \cdots, p_k, \cdots, p_n)_{i_1\cdots j_k \cdots i_n}\nn\\
&&\quad +q_\mu N^\mu (p_1, \cdots, p_n, q=0)_{i_1\cdots i_n a}+{\cal O}(q^2).
\eea
At leading order this leads to
\be
\sum_{k=1}^n (T^a_k)_{i_kj_k}A(p_1\cdots p_k \cdots p_n)_{i_1\cdots j_k\cdots i_n}=0.\label{non}
\ee
Such a relation can be understood as the Ward identity for a global symmetry which is induced from the gauge symmetry by just choosing constant gauge transformation parameter. One can also consider \eqref{non} as the non-Abelian generalization of charge conservation.

At sub-leading order this gives
\be
q_\mu N^\mu (p_1, \cdots, p_n, q=0)_{i_1\cdots i_na}=-\sum_{k=1}^n (T^a_k)_{i_kj_k}q_\mu\frac{\partial}{\partial p_{k\mu}}A(p_1, \cdots, p_k, \cdots, p_n)_{i_1\cdots j_k \cdots i_n}. \ee
As noticed in~\cite{Bern:2014vva}, the solution of $q^\mu E_\mu=0$ which is local in $q^\mu$ can only be
\be
E^\mu=(B_1(p_k, q)\cdot q)B_2^\mu(p_k, q)-(B_2(p_k, q)\cdot q)B_1^\mu(p_k, q).
\ee
Hence it will not contribute to $N^\mu (p_1, \cdots, p_n, q=0)_{i_1\cdots i_na}$. So we have
\be
N^\mu (p_1, \cdots, p_n, q=0)_{i_1\cdots i_na}=-\sum_{k=1}^n (T^a_k)_{i_kj_k}\frac{\partial}{\partial p_{k\mu}}A(p_1, \cdots, p_k, \cdots, p_n)_{i_1\cdots j_k \cdots i_n},
\ee
which leads to
\bea
A^\mu(p_1, \cdots, p_n, q)_{i_1\cdots i_n a}=\sum_{k=1}^n\frac{(T^a_k)_{i_kj_k}}{p_k\cdot q}(p_k^\mu-iq_\nu J^{\mu\nu}_k)A(p_1, \cdots, p_n)_{i_1\cdots i_n}+{\cal O}(q),
\eea
where $J^{\mu\nu}_k$ is the total angular momentum operator. Finally the leading and sub-leading limit of the amplitude are
\begin{align}
A(p_1, \cdots, p_n, q)_{i_1\cdots i_na}&\equiv \epsilon_{q\mu}A^\mu(p_1, \cdots, p_n, q)_{i_1\cdots i_n a}\nn\\
&\to \sum_{k=1}^n (S^{(0)}_k+S^{(1)}_k)(T^a_k)_{i_kj_k}A(p_1, \cdots, p_n)_{i_1\cdots j_k \cdots i_n},
\end{align}
where
\bea
S^{(0)}_k&=&\frac{p_k\cdot \epsilon_q}{p_k\cdot q},\label{leadingfactor}\\
S^{(1)}_k&=&-i\frac{\epsilon_{q\mu}q_\nu J^{\mu\nu}_k}{p_k\cdot q}. \label{subleadingfactor}
\eea

Let us now put the soft factors to asymptotic position space to compare with the result of the Ward identity for asymptotic symmetries. The null momenta need to be parametrized as
\begin{align}
  p_{k\,\mu}&=\frac{\omega_k}{1+w_k\bar{w}_k}\left(1+w_k\bar{w}_k,w_k+\bar{w}_k,i(\bar{w}_k-w_k),1-w_k\bar{w}_k\right),\\
  q_{\mu}&=\frac{\omega_q}{1+w\bar{w}}\left(1+w\bar{w},w+\bar{w},i(\bar{w}-w),1-w\bar{w}\right),
\end{align}
and similarly the polarization tensors as
\begin{equation}
\epsilon_{\mu}^-(q)=\frac{1}{\sqrt{2}}\left(\bar{w},1,-i,-\bar{w}\right),\qquad
\epsilon_{\mu}^+(q)=\frac{1}{\sqrt{2}}\left(w,1,i,-w\right).
\end{equation}
Now we particularize to an outgoing negative (positive)-helicity soft gluon for the leading (sub-leading) soft theorem. The corresponding leading and sub-leading pieces of the soft gluon theorems can be rewritten as
\begin{equation}
\label{soft0+}
	\lim_{\omega_q\to0}\langle\textrm{out}|\omega_q\,\mathfrak{a}^a_+(q)|\textrm{in}\rangle_{i_1 \cdots i_n}=
	\frac{1+|w|^2}{\sqrt{2}}\sum_{k=1}^n \frac{(T^a_k)_{i_kj_k}}{w-w_k}\,\langle\textrm{out}|\textrm{in}\rangle_{i_1\cdots j_k \cdots i_n},
\end{equation}
\begin{multline}
\label{soft1+}
  \lim_{\omega_q\to0}\!\partial_{\omega_q}\langle\textrm{out}|\omega_q\,\mathfrak{a}^a_-(q)|\textrm{in}\rangle_{i_1 \cdots i_n}\!=\! \\
  \sum_{k=1}^n\! \frac{(T^a_k)_{i_kj_k}}{\sqrt{2}}\!\left(\frac{1+w\,\bw_k}{\bw-\bw_k}\partial_{\omega_k}+
  \frac{(1+|w_k|^2)(w-w_k)}{\omega_k(\bw-\bw_k)}\partial_{w_k}\right)\!\langle\textrm{out}|\textrm{in}\rangle_{i_1\cdots j_k \cdots i_n}.
\end{multline}
We have only paid attention to vertices where a gluon couples to two scalars for the sub-leading soft theorem when deriving \eqref{soft1+}, so that the angular momentum operator is just
\be
J^{\mu\nu}_k=i\left(p_k^{\ \mu}\frac{\partial}{\partial p_{k\,\nu}}-p_k^{\ \nu}\frac{\partial}{\partial p_{k\,\mu}}\right),
\ee
without extra helicity terms.


\bibliography{asymptrefs}
\bibliographystyle{utphys}

\end{document}